\documentclass[twocolumn,10pt,superscriptaddress, longbibliography,amsmath,amssymb,aps,pra,floatfix]{revtex4-2}

\usepackage[T1]{fontenc}

\usepackage{graphicx}
\usepackage{qcircuit}
\usepackage{mathtools}
\usepackage{amsthm}
\usepackage{tikz}
\usepackage{color}
\usepackage[normalem]{ulem}  

\usepackage[colorlinks=true,linkcolor=red,citecolor=blue, linktocpage=true,breaklinks=true]{hyperref}

\newcommand{\tr}{\mathrm{Tr}}

\newtheorem{theorem}{Theorem}

\begin{document}

\title{Fluctuation theorems for multipartite quantum coherence and correlation dynamics
}

\author{Kun Zhang}
\affiliation{School of Physics, Northwest University, Xi'an 710127, China}
\affiliation{Shaanxi Key Laboratory for Theoretical Physics Frontiers, Xi'an 710127, China}
\affiliation{Peng Huanwu Center for Fundamental Theory, Xi'an 710127, China}
\affiliation{Fundamental Discipline Research Center for  Quantum Science and technology of Shaanxi Province, Xi'an 710127, China}

\author{Mo-Yang Ni}
\affiliation{School of Physics, Northwest University, Xi'an 710127, China}

\author{Hai-Long Shi}
\email{hailong.shi@ino.cnr.it} 
\affiliation{QSTAR, INO-CNR, and LENS, Largo Enrico Fermi 2, 50125 Firenze, Italy} 

\author{Xiao-Hui Wang}
\email{xhwang@nwu.edu.cn}
\affiliation{School of Physics, Northwest University, Xi'an 710127, China}
\affiliation{Shaanxi Key Laboratory for Theoretical Physics Frontiers, Xi'an 710127, China}
\affiliation{Peng Huanwu Center for Fundamental Theory, Xi'an 710127, China}
\affiliation{Fundamental Discipline Research Center for  Quantum Science and technology of Shaanxi Province, Xi'an 710127, China}

\author{Jin \surname{Wang}}
\email{jin.wang.1@stonybrook.edu}
\affiliation{Department of Chemistry, Stony Brook University, and Department of Physics and Astronomy, Stony Brook University, Stony Brook, New York 11794, USA}

\date{\today}

\begin{abstract}
Fluctuation theorems establish exact relations for nonequilibrium dynamics, profoundly advancing the field of stochastic thermodynamics. In this work, we extend quantum fluctuation theorems beyond the traditional thermodynamic framework to quantum multipartite information dynamics, where both the system and the environment are multipartite without assuming any thermodynamic constraints. Based on the two-point measurement scheme and the classical probability, we establish the fluctuation theorem for the dynamics of classical multipartite mutual information. By extending to quasiprobability, we derive quantum fluctuation theorems for multipartite coherence and quantum correlations, presenting them in both integral and detailed forms. Our theoretical results are illustrated and verified using three-qubit examples, and feasible experimental verification protocols are proposed. These findings uncover the statistical structure underlying the nonequilibrium quantum information dynamics, providing fundamental insights and alternative tools for advancing quantum technologies. 

\end{abstract}

\maketitle

\section{Introduction}

Quantum coherence and correlations in many-body systems encode the essential nonclassical features of quantum matter and are indispensable resources for quantum technologies~\cite{laurell2025witnessing,RevModPhys.84.1655,RevModPhys.89.041003,RevModPhys.90.035005,DeChiara2017GenuineQC}. Understanding the fundamental limits governing the dynamics of these quantum resources is therefore crucial for advancing information processing in physical systems. In particular, information inequalities set fundamental bounds on the evolution of coherence and correlations~\cite{Wilde2011FromCT}.

On the other hand, thermodynamic inequalities, such as the second law of thermodynamics, highlight the inherent constraints within the fundamental laws of physics \cite{landiIrreversibleEntropyProduction2021}. Beyond the macroscopic constraint imposed by the second law, the positivity of entropy production allows for a more refined characterization through an exact equality known as the fluctuation theorem~\cite{Evans1993ProbabilityOS,jarzynskiNonequilibriumEqualityFree1997,Crooks1999EntropyPF,Evans2002,Jarzynski2011EqualitiesAI}. This theorem has fundamentally reshaped our understanding of thermodynamics by connecting microscopic dynamics to macroscopic laws, and it has become a cornerstone of stochastic thermodynamics, offering detailed statistical descriptions of nonequilibrium processes \cite{Seifert2007StochasticTP,Seifert2012StochasticTF}.

Existing fluctuation theorems are primarily grounded in the framework of thermodynamics, typically focusing on bipartite system-environment models \cite{vanZon2003ExtensionOT,Jarzynski2004ClassicalAQ,Seifert2005EntropyPA,Douarche2006WorkFT,Esposito2009ThreeDF,Noh2012Fluctuation,Michel2012LocalFT,jevtic2015exchange,santos2020joint,Pei2023PromotingFT}, or extending to tripartite settings that incorporate Maxwell's demon \cite{Kim2007FluctuationTF,Sagawa2009GeneralizedJE,Ponmurugan2010GeneralizedDF,Horowitz2010NonequilibriumDF,Sagawa2011NonequilibriumTO,Lahiri2011FluctuationTI,Sagawa2012FluctuationTW,Kundu2012NonequilibriumFT,Zeng2021NewFT}. 
Quantum generalizations have been widely explored \cite{tasakiJarzynskiRelationsQuantum2000,Sagawa2007SecondLO,Deffner2011NonequilibriumEP,Iyoda2016FluctuationTF,Lostaglio2017QuantumFT,2016Fully,Bartolotta2017JarzynskiEF,Manzano2017QuantumFT,Morris2018QuantumCF,Yada2021QuantumFT,Zhang2021ConditionalEP,DeChiara2022QuantumFT,Wu2024GeneralizedQF,Prech2023QuantumFT,Beyer2024OperationalWF,Esposito2008NonequilibriumFF,campisiColloquiumQuantumFluctuation2011}, yet they remain largely constrained to thermodynamic observables, with information typically treated as an additional term.
In contrast, recent advances in quantum many-body dynamics have revealed that diverse nonequilibrium phenomena, such as operator spreading, information scrambling, and entanglement growth, go beyond the scope of conventional thermodynamic descriptions \cite{aolita2015open,PhysRevB.95.094302,PhysRevX.7.031016,PhysRevX.8.021013,PhysRevX.8.021014,lewis2019dynamics,PhysRevX.9.041017,PhysRevD.106.046007,PhysRevLett.130.250401}.
This motivates the pursuit of fluctuation theorems that probe quantum information dynamics beyond the traditional thermodynamic scope.

Most known quantum fluctuation theorems rely on the two-point measurement scheme \cite{tasakiJarzynskiRelationsQuantum2000}. 
The quantum-to-classical transition induced by projective measurements poses a fundamental challenge to formulating fluctuation theorems of quantum coherence and correlations. 
In recent years, alternative approaches, such as Bayesian networks \cite{Micadei2019QuantumFT,Park2020InformationFT} and single-point measurement schemes \cite{Deffner2016QuantumWA,Sone2020QuantumJE,Sone2022JarzynskilikeEO,Sone2022ExchangeFT,Maeda2023DetailedFT}, have been proposed, offering new possibilities for extending the application of quantum fluctuation theorems.
This naturally raises an intriguing question: can information inequalities be refined at the microscopic level in the form of fluctuation theorems, especially for many-body quantum information? 
Such an extension would not only deepen our understanding of information dynamics but also establish a universal theoretical framework for analyzing information-driven protocols.

\begin{figure}[t]
\includegraphics[width=1\columnwidth]{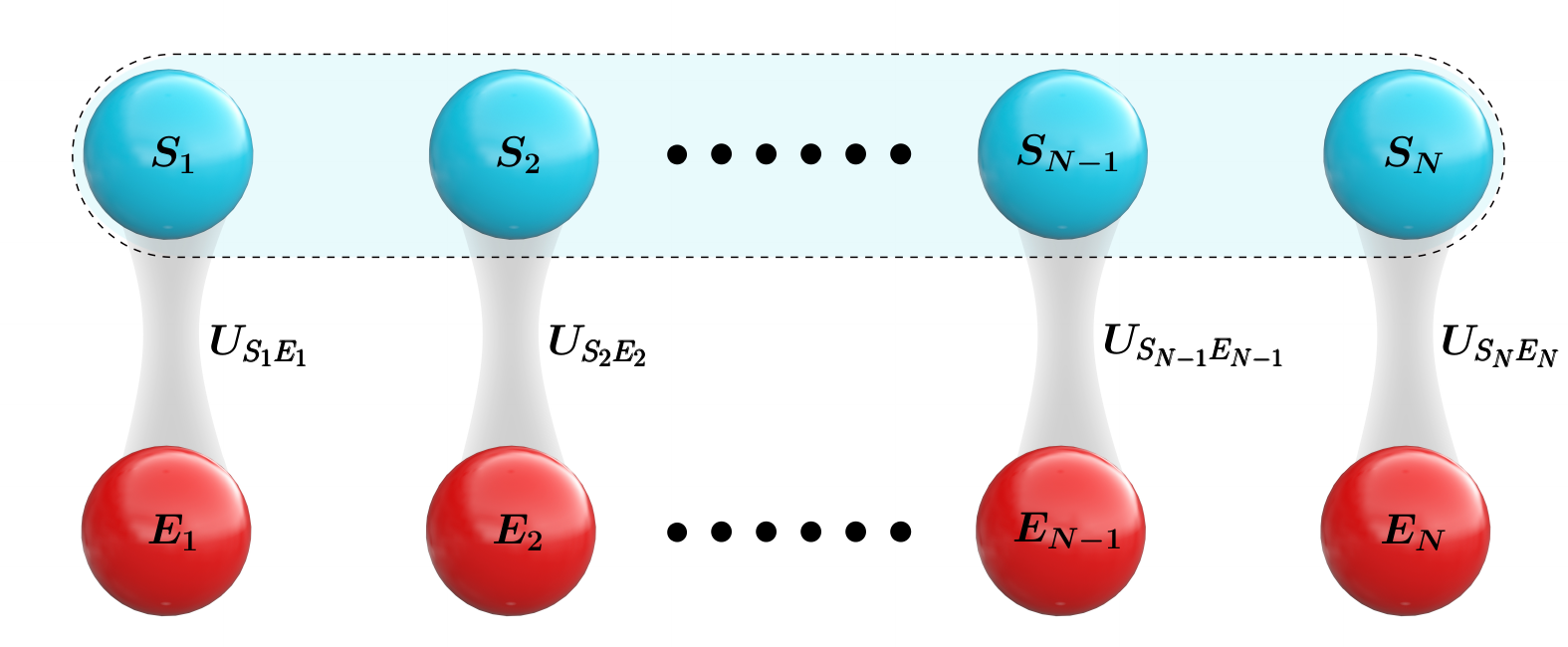}
\caption{Schematic of the dynamics of an $N$-partite correlated state $\rho_S = \rho_{S_1S_2\cdots S_N}$ interacting with multipartite environments $\rho_E = \bigotimes_{j=1}^N\rho_{E_j}$. Initially, the subenvironments are uncorrelated. The interaction between the subsystem $S_j$ and the subenvironment $E_j$ is denoted as $U_{S_jE_j}$, where $j = 1,2,\cdots, N$.}
\label{fig_model}
\end{figure}

In this paper, we establish quantum fluctuation theorems for multipartite classical correlations, quantum coherence, and quantum correlations within a unified framework, considering a system of $N$ subsystems each interacting with individual subenvironments (see Fig.~\ref{fig_model}). 
By applying the two-point measurement scheme, we formulate the fluctuation theorem for many-body classical correlations via the classical probability.
To capture non-classical features, we extend the above classical probability framework to the quasiprobability approach, an indispensable tool in quantum information science \cite{Ferrie2010QuasiprobabilityRO,Gherardini2024QuasiprobabilitiesIQ,ArvidssonShukur2024PropertiesAA}.
In recent years, this approach has also found growing applications in the study of quantum fluctuation theorems \cite{Halpern2016JarzynskilikeEF,Halpern2017QuasiprobabilityBT,Kwon2018FluctuationTF,Levy2019QuasiprobabilityDF,Huang2022FluctuationTF,Zhang2022QuasiprobabilityFT}. 
Our results demonstrate that the quasiprobability framework faithfully captures the statistics of both quantum coherence and correlations, thereby offering a unified perspective on the dynamics of quantum resources.



This paper is organized as follows. In Sec. \ref{sec:setup}, we set up the model and describe its information dynamics. The fluctuation theorem for the total classical correlation, based on the two-point measurement scheme, is presented in Sec. \ref{sec:classical_correlation}. We then establish quasiprobability fluctuation theorems for the total quantum correlation and for quantum coherence in Secs. \ref{sec:quantum_correlation} and \ref{sec:quantum_coherence}, respectively. Examples based on three-qubit dynamical models are provided in Sec. \ref{sec:example}. Finally, conclusions and discussions including the experimental verification proposals are given in Sec. \ref{sec:conclusion}.

\section{Dynamics of multipartite coherence and correlation}

\label{sec:setup}

We start from an initial $N$-partite state $\rho_S= \rho_{S_1S_2\cdots S_N}$, which contains the many-body coherence and correlations.
Each subsystem $S_j$ locally interacts with an environment $E_j$ (see Fig.~\ref{fig_model}), with the total environment initially uncorrelated across all sites, i.e., $\rho_E = \bigotimes_{j=1}^N\rho_{E_j}$.
The local interactions are described by unitary operators  $U_{S_jE_j}$, yielding the global evolution $U_{SE} = \bigotimes_{j=1}^NU_{S_jE_j}$.
The resulting state after evolution is a $2N$-partite correlated state 
\begin{equation}
    \rho'_{SE} = U_{SE}(\rho_S\otimes \rho_E)U^\dag_{SE},
\end{equation}
in which the quantum resources initially present in $\rho_S$ are generally dissipated. 
Throughout, we use a prime to denote quantities associated with the final state. Because of the presence of multipartite correlation in $\rho_S$, the $N$-particle dynamics is not equivalent to $N$ copies of single-particle dynamics.

We quantify the many-body quantum correlation by the total quantum mutual information, also known as the total quantum correlation, defined as \cite{Modi2009UnifiedVO,DeChiara2017GenuineQC}
\begin{equation}
    \mathcal I(\rho_S) = \sum_{j=1}^N \mathcal S(\rho_{S_j})-\mathcal S(\rho_{S}),
\end{equation}
where $\mathcal S(\rho) = -\tr(\rho\ln \rho)$ is the von Neumann entropy.
This generalizes the widely used bipartite mutual information and measures the closest distance between $\rho_S$ and a product state.
It serves as a useful probe for many-body localization \cite{Goold2015TotalCO,Pietracaprina2016TotalCO} and local relaxation \cite{Li2020TheHR,Li2024RelaxingTG}. 


The change in the total quantum correlation is denoted as \footnote{We define the total quantum correlation change as the initial value minus its final value. Consequently, it is non-negative, analogous to the positivity of entropy production in the second law of thermodynamics.}
\begin{equation}
    \Delta \mathcal{I} = \mathcal{I}(\rho_S) - \mathcal{I}(\rho'_S),
\end{equation}
where $\rho'_S$ is the evolved state under the dynamics $U_{SE}$. 
Intuitively, the correlations can only decrease, i.e., $\Delta\mathcal I\geq 0$, because $U_{SE}$ acts locally on each subsystem. This is rigorously guaranteed by the strong subadditivity of the von Neumann entropy \cite{liebProofStrongSubadditivity1973}. 
For $N=2$, this reduces to the well-known quantum data-processing inequality \cite{schumacherQuantumDataProcessing1996}. 
Note that $\Delta \mathcal{I} \geq 0$ holds as long as the initial environment is a multipartite product state \cite{buscemi2014complete}.

Strictly speaking, the mutual information $\mathcal I(\rho_S)$ includes both classical and quantum correlations. The amount of classical correlation in $\rho_S$ can be quantified as
\begin{equation}
    \mathcal I_\text{cl}(\rho_S) = \mathcal I(\tilde\rho_S),
\end{equation}
where the classical state $\tilde{\rho}_{S}$ is obtained by the decoherence operation \cite{Luo2008UsingMD,Modi2009UnifiedVO}
\begin{equation}
    \tilde{\rho}_{S} = \sum_{\{s\}}\Pi_{\{s\}}\rho_{S}\Pi_{\{s\}}.
\end{equation}
Here $\Pi_{\{s\}} = \bigotimes_{j=1}^N \Pi_{s_j}$ is the product of local rank-1 eigenprojectors $\Pi_{s_j} = |s_j\rangle\langle s_j|$ of $\rho_{S_j}$ and $\{s\} = \{s_1,s_2,\cdots,s_N\}$. Note that we have the spectral decomposition $\rho_{S_j} = \sum_{s_j}p_{s_j}|s_j\rangle\langle s_j|$. The classical state $\tilde{\rho}_{S}$ is obtained from $\rho_S$ by local projective measurements, therefore the entropy can only increase, namely $\mathcal S(\tilde\rho_S)\geq \mathcal S(\rho_S)$ \cite{nielsenQuantumComputationQuantum2010}. However, the local state is not affected by the local eigenprojectors, namely $\tilde\rho_{S_j} = \rho_{S_j}$. Consequently, we have the inequality $\mathcal I(\rho_S)\geq \mathcal I(\tilde\rho_S)$ with the equal sign if $\rho_S = \tilde\rho_S$.


We then denote the change in the total classical correlation as 
\begin{equation}
    \Delta\mathcal I_\text{cl} = \mathcal I_\text{cl}(\rho_S) - \mathcal I_\text{cl}(\tilde\rho'_S),
\end{equation}
where $\tilde\rho'_S = \tr_E\tilde\rho'_{SE}$ and $\tilde\rho'_{SE} = U_{SE}(\tilde\rho_S\otimes \rho_E)U^\dag_{SE}$. A subtle point is that we use the initial state $\tilde\rho_S$ for the evolution, rather than $\rho_S$. In this way, the dynamics can be modeled as a classical stochastic process by tracking the probabilities of states on the local eigenbasis. Consequently, the information inequality $\Delta\mathcal I_\text{cl}\geq 0$ can be established, which is a multipartite version of the classical data-processing inequality \cite{cover1999elements}.

The difference between $\mathcal I(\rho_S)$ and $\mathcal I_\text{cl}(\rho_S)$ reflects correlations arising from quantum phenomena. It can also be interpreted as the amount of many-body coherence in $\rho_S$, denoted as
\begin{equation}
    \mathcal C(\rho_S) = \mathcal I(\rho_S) - \mathcal I_\text{cl}(\rho_S),
\end{equation}
with respect to the local eigenbasis \cite{Baumgratz2013QuantifyingC,Yao2015QuantumCI}. We denote the change in many-body coherence as 
\begin{equation}
    \Delta\mathcal C = \mathcal C(\rho_S) - \mathcal C(\rho'_S).
\end{equation}
The coherence change $\Delta\mathcal C$ is not necessarily positive for arbitrary evolutions $U_{SE}$. However, if the dynamics results in a final state which is completely dephased, satisfying 
\begin{equation}
\label{eq:dephasing_condition}
    \rho'_S = \sum_{s'}\Pi_{\{s'\}}\rho_S'\Pi_{\{s'\}},
\end{equation}
then the final state has $\mathcal C(\rho'_S) = 0$. Consequently, we always have $\Delta\mathcal C\geq 0$. Furthermore, the total quantum correlation change admits the decomposition 
\begin{equation}
\label{eq:Delta_I_decomposition}
    \Delta\mathcal I = \Delta\mathcal I_\text{cl} + \Delta\mathcal C,
\end{equation}
separating the classical and quantum contributions.

The information inequalities $\Delta\mathcal I,\Delta\mathcal I_\text{cl},\Delta\mathcal C\geq 0$ describe the unidirectional flow of many-body correlations and coherence from the system to the environment, akin to the thermodynamic flow of heat from hot to cold. Such irreversible thermodynamic processes are quantitatively characterized by fluctuation theorems \cite{Jarzynski2004ClassicalAQ,Noh2012Fluctuation,Levy2019QuasiprobabilityDF,Wu2024GeneralizedQF}. Next, we show that each of the quantities $\Delta\mathcal I$, $\Delta\mathcal I_\text{cl}$, and $\Delta\mathcal C$ obeys a fluctuation theorem.

\section{Fluctuation theorem of total classical correlation dynamics}

\label{sec:classical_correlation}

%
%
The von Neumann entropy of the initial classical state is given by 
\begin{equation}
    \mathcal S(\tilde\rho_S) = -\sum_{{s}} p_{{s}} \ln p_{{s}},
\end{equation}
where the probability distribution $p_{{s}}$ is the eigenvalues of $\tilde\rho_S$. In other words, we have
\begin{equation}
    \tilde\rho_S = \sum_{{s}} p_{{s}} \Pi_{{s}}.
\end{equation}
Accordingly, the quantity $(-\ln p_{{s}})$ can be identified as a stochastic entropy variable, whose average over $p_{{s}}$ yields $\mathcal S(\tilde\rho_S)$.
Analogously, the stochastic mutual information is defined as 
\begin{equation}
    \iota(\tilde\rho_S) = \ln p_{{s}} - \sum_{j=1}^N \ln p_{s_j}, 
\end{equation}
recovering the mutual information $\mathcal I(\tilde\rho_S)$ upon averaging over $p_{{s}}$.
Here, $p_{s_j}$ is the eigen-probability of the reduced density matrix of the subsystem $S_j$, which has the decomposition
\begin{equation}
    \rho_{S_j} = \sum_{s_j} p_{s_j} \Pi_{s_j}.
\end{equation}
The stochastic mutual information is the key when considering the fluctuation theorems involving the measurement and feedback \cite{Sagawa2009GeneralizedJE,Ponmurugan2010GeneralizedDF,Horowitz2010NonequilibriumDF,Sagawa2011NonequilibriumTO,Lahiri2011FluctuationTI,Sagawa2012FluctuationTW}.
Corresponding to the total classical correlation change $\Delta\mathcal I_\text{cl}$, we define the stochastic total classical correlation change as
\begin{equation}
\label{eq:delta_iota_cl}
    \Delta\iota_\text{cl} = \ln p_{\{s\}} - \sum_{j=1}^N \ln p_{s_j} - \left(\ln p_{\{\tilde s'\}} - \sum_{l=1}^N \ln p_{\tilde s'_l}\right),
\end{equation}
where $p_{\{\tilde s'\}}$ and $p_{\tilde s'_j}$ are eigen-probabilities of the final states $\tilde\rho'_S$ and $\tilde\rho'_{S_j}$, respectively. 


Since $\Delta\iota_\text{cl}$ includes both $p_{\{s\}}$ and $p_{\{\tilde s'\}}$, which correspond to the initial and final states, respectively, it is natural to consider the joint probability of the initial and final states,  given by
\begin{equation}
\label{eq:P_F}
    \mathcal P^{\text F}[\xi] = \tr\left(U^\dag_{SE}\Pi_{\{\tilde s'\}}\Pi_{\{\tilde n'\}}U_{SE}\Pi_{\{s\}}\Pi_{\{n\}}\tilde\rho_S\rho_E\right),
\end{equation}
with the stochastic index $\xi = \{s,n,\tilde s',\tilde n'\}$. Here $\Pi_{\{s\}}$ ($\Pi_{\{\tilde s'\}}$) and $\Pi_{\{n\}}$ ($\Pi_{\{\tilde n'\}}$) are local eigenprojectors of the initial (final) states of system and environment, respectively. For simplicity, we omit the tensor product symbol, e.g., $\tilde\rho_S\rho_E = \tilde\rho_S\otimes\rho_E$. The projectors $\Pi_{\{s\}}$ and $\Pi_{\{n\}}$ are non-invasive with respect to $\tilde\rho_S$ and $\rho_E$, ensuring that $\mathcal P^{\text F}$ is a valid classical probability. This construction corresponds to the two-point measurement scheme commonly employed in the study of quantum fluctuation theorems \cite{Esposito2008NonequilibriumFF}.

By combining the stochastic total classical correlation change $\Delta\iota_\text{cl}$ defined in Eq. (\ref{eq:delta_iota_cl}) and the joint probability distribution $\mathcal P^{\text F}$ defined in Eq. (\ref{eq:P_F}), we obtain 
\begin{equation}
    \langle \Delta\iota_\text{cl}\rangle_{\mathcal P^{\text F}[\xi]} = \Delta\mathcal I_\text{cl},
\end{equation}
with the notation $\langle\cdot\rangle_{\mathcal P^{\text F}[\xi]} = \sum_\xi \mathcal P^{\text F}[\xi](\cdot)$. This identifies $\langle \Delta\iota_\text{cl}\rangle_{\mathcal P^{\text F}[\xi]}$ as the statistical version of $\Delta\mathcal I_\text{cl}$, enabling the study of higher-order statistical moments. Furthermore, the fluctuation theorem can be established as follows. 
\begin{theorem}
The stochastic total classical correlation change $\Delta\iota_\mathrm{cl}$ satisfies the integral fluctuation relation
    \begin{equation}
    \label{eq:FT_i_cl}
        \langle e^{-\Delta\iota_\mathrm{cl}} \rangle_{\mathcal P^{\mathrm F}[\xi]} = 1.
    \end{equation}
\end{theorem}

\begin{proof}
First, by the definition in Eq. (\ref{eq:delta_iota_cl}), the stochastic total classical correlation change $\Delta\iota_\mathrm{cl}$ leads to
    \begin{equation}
        e^{-\Delta\iota_\text{cl}} = \frac{p_{\{\tilde s'\}}\prod_{j=1}^N p_{s_j}}{p_{\{s\}}\prod_{l=1}^N p_{\tilde s'_l}}.
    \end{equation}
    Then the direct calculation gives
    \begin{widetext}
    \begin{align}
        \langle e^{-\Delta\iota_\text{cl}} \rangle_{\mathcal P^{\text F}[\xi]} = & \sum_\xi\frac{p_{\{\tilde s'\}}\prod_{j=1}^N p_{s_j}}{p_{\{s\}}\prod_{l=1}^N p_{\tilde s'_l}}\tr\left(U^\dag_{SE}\Pi_{\{\tilde s'\}}\Pi_{\{\tilde n'\}}U_{SE}\Pi_{\{s\}}\Pi_{\{n\}}\tilde\rho_S\rho_E\right) \nonumber \\
        = & \sum_\xi\frac{p_{\{\tilde s'\}}\prod_{j=1}^N p_{s_j}}{p_{\{s\}}\prod_{l=1}^N p_{\tilde s'_l}} \tr\left(U^\dag_{SE}\Pi_{\{\tilde s'\}}\Pi_{\{\tilde n'\}}U_{SE}\Pi_{\{s\}}\Pi_{\{n\}}\right)p_{\{s\}}\prod_{m=1}^N p_{n_m} \nonumber \\
        = & \sum_\xi\frac{p_{\{\tilde s'\}}}{\prod_{l=1}^N p_{\tilde s'_l}} \tr\left(U^\dag_{SE}\Pi_{\{\tilde s'\}}\Pi_{\{\tilde n'\}}U_{SE}\Pi_{\{s\}}\Pi_{\{n\}}\right) \prod_{j=1}^N p_{s_j}\prod_{m=1}^N p_{n_m} \nonumber \\
        = & \sum_{\tilde s',\tilde n'}\frac{p_{\{\tilde s'\}}}{\prod_{l=1}^N p_{\tilde s'_l}}\tr\left(U^\dag_{SE}\Pi_{\{\tilde s'\}}\Pi_{\{\tilde n'\}}U_{SE}\bigotimes_{j=1}^N\rho_{S_j}\rho_E\right) \nonumber \\
        = & \sum_{\tilde s',\tilde n'}\frac{p_{\{\tilde s'\}}}{\prod_{l=1}^N p_{\tilde s'_l}} \prod_{j=1}^N p_{\tilde s'_j} \prod_{m=1}^N p_{\tilde n'_m} \nonumber \\
        = & \sum_{\tilde s',\tilde n'}p_{\{\tilde s'\}}\prod_{m=1}^N p_{\tilde n'_m} \nonumber \\
        = & 1.
    \end{align}
    \end{widetext}
    The first to the second line is from the eigen-decompositions of $\tilde\rho_S$ and $\rho_E$, denoted as
        \begin{equation}
        \tilde\rho_S = \sum_s p_{\{s\}}\Pi_{\{s\}},\qquad \rho_E = \sum_n p_{\{n\}}\Pi_{\{n\}},
    \end{equation}
    with $p_{\{n\}} = \prod_{m=1}^N p_{n_m}$. The second to the third line is to cancel $p_{\{s\}}$. The third to the fourth line is due to $\rho_{S_j} = \sum_{s_j}p_{s_j}\Pi_{s_j}$. The fourth to the fifth line is by applying the relation
    \begin{equation}
    \tr\left(U^\dag_{S_jE_j}\Pi_{\tilde s'_j}\Pi_{\tilde n'_j}U_{S_jE_j}\rho_{S_j}\rho_{E_j}\right) = p_{\tilde s'_j}p_{\tilde n'_j}.
    \end{equation}
\end{proof}

Applying Jensen's inequality, $e^{-\langle x\rangle}\leq \langle e^{-x}\rangle$, the fluctuation theorem of the total classical correlation immediately yields the information inequality $\Delta\mathcal I_\text{cl}\geq 0$. Expanding Eq. (\ref{eq:FT_i_cl}) further gives 
\begin{equation}
    \langle \Delta\iota_\text{cl}^2\rangle_{\mathcal P^{\text F}[\xi]} = 2\Delta\mathcal I_\text{cl} + \mathcal O(\Delta\iota_\text{cl}^3),
\end{equation}
establishing a direct connection between the expectation value and the second moment distribution. Additional constraints on higher-order moments can also be derived from the fluctuation theorem \cite{merhavStatisticalPropertiesEntropy2010}. 

Alternatively, the fluctuation theorem can be understood as a symmetry between the physical forward evolution and its reversed process, leading to Crooks fluctuation relation \cite{Crooks1999EntropyPF} or the detailed fluctuation theorem \cite{Esposito2008NonequilibriumFF}. The reversed process serves as a retrodiction of the initial state from the final state \cite{watanabe1955symmetry,buscemi2021fluctuation,aw2021fluctuation}. In our case, the joint probability of the reversed process is defined as
\begin{equation}
\label{eq:P_B}
    \mathcal P^{\text B}[\xi] = \tr\left(U_{SE}\Pi_{\{s\}}\Pi_{\{n\}}U^\dag_{SE}\Pi_{\{\tilde s'\}}\Pi_{\{\tilde n'\}}\tilde\rho^\text{ref}_S\tilde\rho^\text{ref}_E\right),
\end{equation}
where $\tilde\rho^\text{ref}_S$ is the system's reference state given by
\begin{equation}
    \tilde\rho^\text{ref}_S = \sum_{\tilde s'}p_{\{\tilde s'\}}\Pi_{\{\tilde s'\}},
\end{equation}
and $\tilde\rho^\text{ref}_E$ is the environmental reference state given by
\begin{equation}
    \tilde\rho^\text{ref}_E = \bigotimes_{j=1}^N \tilde\rho'_{E_j}
\end{equation}
where $\tilde\rho'_{E_j}$ is the final state of the $j$-th subenvironment, obtained from $\tilde\rho'_{E} = \tr_S\tilde\rho'_{SE}$. Note that the reference state $\tilde\rho^\text{ref}_S$ satisfies the completely dephased condition given in Eq. (\ref{eq:dephasing_condition}).

To analyze the local dynamics, we marginalize over variables unrelated to the subsystem $S_j$ and the subenvironment $E_j$, yielding the local joint probabilities 
\begin{equation}
    \sum_{\xi/\xi_j}\mathcal P^{\text{F(B)}}[\xi] = \mathcal P_{j}^{\text{F(B)}}[\xi_j],
\end{equation}
with $\xi_j = \{s_j,n_j,\tilde s'_j,\tilde n'_j\}$. 
\begin{theorem}
    The stochastic total classical correlation change $\Delta\iota_\mathrm{cl}$ satisfies the detailed fluctuation relation
    \begin{equation}
    \label{eq:dFT_i_cl}
        \frac{\mathcal P^\mathrm{F}[\xi]}{\prod_{j=1}^N \mathcal P_{j}^\mathrm{F}[\xi_j]} e^{-\Delta\iota_\mathrm{cl}} = \frac{\mathcal P^\mathrm{B}[\xi]}{\prod_{j=1}^N \mathcal P_{j}^\mathrm{B}[\xi_j]}.
    \end{equation}
\end{theorem}
\begin{proof}
    The forward process probability $\mathcal P^\text{F}$ defined in Eq. (\ref{eq:P_F}) can be rewritten as
    \begin{equation}
    \mathcal P^{\text F}[\xi] = \prod_{l=1}^N |\langle \tilde s'_l\tilde n'_l|U_{S_lE_l}|s_ln_l\rangle|^2  p_{\{s\}} \prod_{m=1}^N p_{n_m}.
    \end{equation}
    Its marginalization gives the probabilities of local dynamics, namely
    \begin{equation}
        \mathcal P_{j}^{\text F}[\xi_j] = \sum_{\xi/\xi_j}\mathcal P^{\text F}[\xi] =  |\langle \tilde s_j'\tilde n_j'|U_{S_jE_j}|s_jn_j\rangle|^2p_{s_j}p_{n_j}.
    \end{equation}
    Then we have the relation
    \begin{equation}
    \label{eq:P_F_ratio}
        \frac{\mathcal P^\text{F}[\xi]}{\prod_{j=1}^N\mathcal P_j^\text{F}[\xi_j]} = \frac{p_{\{s\}}}{\prod_{j=1}^N p_{s_j}}.
    \end{equation}
    The probability of the reversed process in Eq. (\ref{eq:P_B}) can be rewritten as
    \begin{equation}
        \mathcal P^{\text B}[\xi] = \prod_{l=1}^N |\langle s_ln_l|U^\dag_{S_lE_l}|\tilde s'_l\tilde n'_l\rangle|^2 p_{\{\tilde s'\}} \prod_{m=1}^N p_{\tilde n'_m}.
    \end{equation}
    The local dynamics has
    \begin{equation}
        \mathcal P_{j}^{\text B}[\xi_j] = \sum_{\xi/\xi_j}\mathcal P^{\text B}[\xi] = |\langle s_jn_j|U^\dag_{S_jE_j}|\tilde s'_j\tilde n'_j\rangle|^2 p_{\tilde s'_j}p_{\tilde n'_j}.
    \end{equation}
    Therefore, we have
    \begin{equation}
    \label{eq:P_B_ratio}
        \frac{\mathcal P^\text{B}[\xi]}{\prod_{j=1}^N\mathcal P_j^\text{B}[\xi_j]} = \frac{p_{\{\tilde s'\}}}{\prod_{j=1}^N p_{\tilde s'_j}}.
    \end{equation}
    Combining the relations (\ref{eq:P_F_ratio}) and (\ref{eq:P_B_ratio}), we can prove
    \begin{align}
        \frac{\mathcal P^{\text F}[\xi]}{\prod_{l=1}^N \mathcal P_{l}^{\text F}[\xi_l]} e^{-\Delta\iota_\text{cl}} = & \frac{p_{\{s\}}}{\prod_{l=1}^N p_{s_l}}\frac{p_{\{\tilde s'\}}\prod_{m=1}^N p_{s_m}}{p_{\{s\}}\prod_{j=1}^N p_{\tilde s'_j}} \nonumber \\
        = & \frac{p_{\{\tilde s'\}}}{\prod_{j=1}^N p_{\tilde s'_j}} \nonumber \\
        = & \frac{\mathcal P^{\text B}[\xi]}{\prod_{j=1}^N \mathcal P_{j}^{\text B}[\xi_j]}.
    \end{align}
\end{proof}



The detailed fluctuation relation in Eq.~(\ref{eq:dFT_i_cl}) is fundamentally distinct from conventional fluctuation theorems for the entropy production, whether at the level of the total system $S$ or its subsystems $S_j$ \cite{landiIrreversibleEntropyProduction2021}. 
Notably, it is constructed from conditional probabilities linking the joint and local distributions, $\mathcal P^\mathrm{F}$ and $\mathcal P_j^\mathrm{F}$. It also differs from classical fluctuation theorems formulated in causal networks, where information appears as an additional correction term modifying the second law \cite{Ito2013InformationTO,Wolpert2019UncertaintyRA}.

\begin{figure*}[t]
\includegraphics[width=1\textwidth]{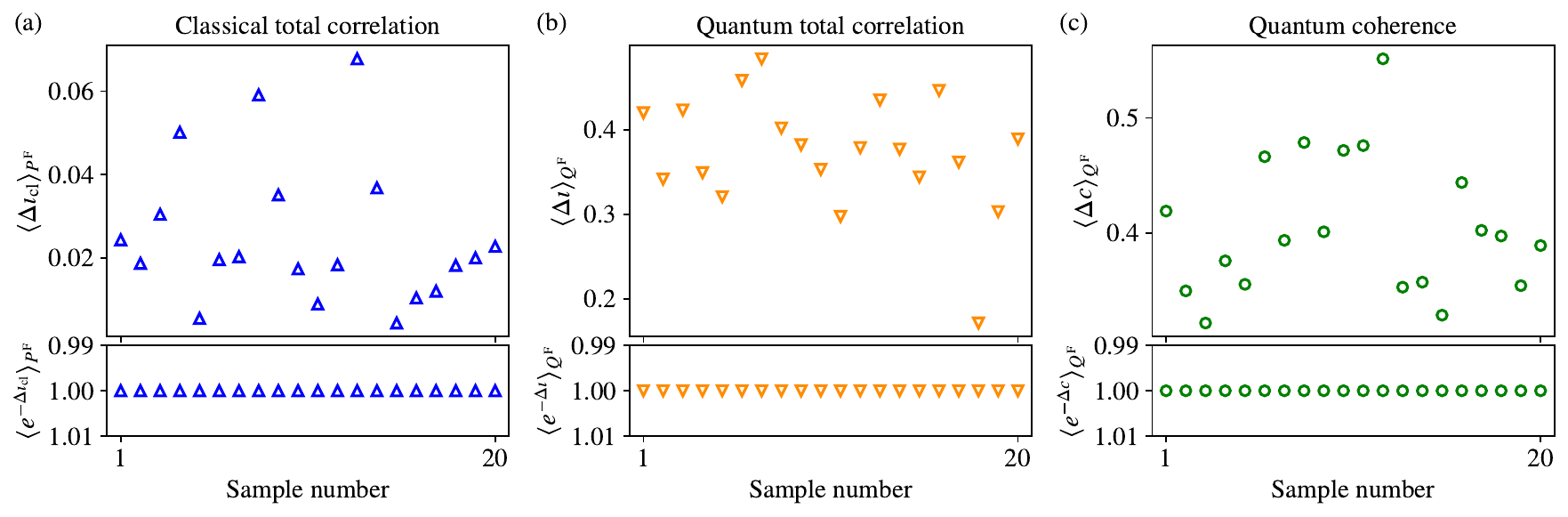}
\caption{Verifications of the integral fluctuation relations (\ref{eq:FT_i_cl}), (\ref{eq:FT_i_q}), and (\ref{eq:FT_c}), through three-qubit examples. The initial states of the system, $\rho_S$ or $\tilde\rho_S$, and the environment qubits $\rho_{E_j}$ are randomly generated. For the total classical and quantum correlations, the dynamics $U_{SE}$ are randomly chosen. For the coherence change, the dynamics is set to the two-qubit swap gate, namely $U_{S_jE_j} = \text{SWAP}$.}
\label{fig_ft}
\end{figure*}

\section{Fluctuation theorem of total quantum correlation dynamics}

\label{sec:quantum_correlation}

Having established quantum fluctuation theorems for classical correlations in many-body systems, we now extend the formulation to encompass quantum correlations. Based on the definition of the total quantum correlation $\mathcal I(\rho_S)$ and the corresponding dynamical change $\Delta\mathcal I$, we define the stochastic total quantum correlation change as
\begin{equation}
\label{eq:delta_iota_q}
    \Delta\iota = \ln p_{k} - \sum_{j=1}^N\ln p_{s_j} - \left(\ln p_{k'} - \sum_{l=1}^N\ln p_{s'_l}\right),
\end{equation}
where the probabilities $p_k$ and $p_k'$ are obtained from the spectral decompositions 
\begin{equation}
    \rho_S = \sum_kp_k\Pi_k,\qquad \rho'_S=\sum_{k'}p_{k'}\Pi_{k'},
\end{equation}
while the probabilities $p_{s_j}$ and $p_{s'_j}$ are obtained from 
\begin{equation}
    \rho_{S_j} = \sum_{s_j}p_{s_j}\Pi_{s_j},\qquad \rho'_{S_j} = \sum_{s_j}p_{s'_j}\Pi_{s'_j}.
\end{equation}
Note that the stochastic mutual information follows the fluctuation theorem based on the Bayesian networks method \cite{Micadei2019QuantumFT}. However, we aim to address the fluctuation theorem of the mutual information change. 


We aim to identify a probability distribution with marginals $p_k$ ($p_{k'}$) and $p_{s_j}$ ($p_{s'_j}$), such that the expectation value of $\Delta\iota$ recovers $\Delta\mathcal I$.
However, since the joint eigenprojector $\Pi_k$ of quantum-correlated states generally fails to commute with the local eigenprojector $\Pi_{s_j}$,  a joint probability distribution for $p_k$ and $p_{s_j}$ cannot be properly defined.
This reflects the quantum contextuality \cite{Lostaglio2017QuantumFT}.
To overcome this, we generalize the classical probability description in Eq. \eqref{eq:P_F} to a quasiprobability by introducing
\begin{equation}
    \label{def:Q_F}
        \mathcal Q^\text{F}[\zeta] = \tr\left(U_{SE}^\dag \Pi_{k'}\Pi_{\{s'\}}\Pi_{\{n'\}}U_{SE}\Pi_{\{s\}}\Pi_{\{n\}}\Pi_{k}\rho_S\rho_E\right),
    \end{equation}
with $\zeta = \{k,s,n,k',s',n'\}$.  
While quasiprobability $\mathcal Q^\text{F}$ is properly normalized, namely
\begin{equation}
    \sum_\zeta \mathcal Q^\text{F}[\zeta] = 1,
\end{equation}
it can take negative or complex values, featuring the statistics of noncommuting observables.  Notably, when $\rho_S$ is classical, the projectors $\Pi_k$ and $\Pi_{{s}}$ coincide, reducing $\mathcal Q^\text{F}$ to the two-point measurement probability $\mathcal P^\text{F}$ defined in Eq.~(\ref{eq:P_F}).

It can be verified that the expectation value of $\Delta\iota$ over $\mathcal Q^\text{F}$ correctly yields 
\begin{equation}
    \langle \Delta\iota\rangle_{\mathcal Q^{\text F}[\zeta]} = \Delta\mathcal I.
\end{equation}
Moreover, the statistics of $\Delta\iota$ are governed by the following fluctuation theorem.
\begin{theorem}
    The stochastic total quantum correlation change $\Delta\iota$ satisfies the integral fluctuation relation
    \begin{equation}
    \label{eq:FT_i_q}
        \langle e^{-\Delta\iota} \rangle_{\mathcal Q^\mathrm{F}[\zeta]} = 1.
    \end{equation}
\end{theorem}
\begin{proof}
    By direct calculation, we have
    \begin{widetext}
    \begin{align}
    \label{proof:FT_q_mutual_information}
        \langle e^{-\Delta\iota} \rangle_{\mathcal Q^{\text F}[\zeta]} = & \sum_\zeta\frac{p_{k'}\prod_{j=1}^N p_{s_j}}{p_{k}\prod_{l=1}^N p_{s'_l}} \tr\left(U_{SE}^\dag \Pi_{k'}\Pi_{\{s'\}}\Pi_{\{n'\}}U_{SE}\Pi_{\{s\}}\Pi_{\{n\}}\Pi_{k}\rho_S\rho_E\right) \nonumber \\
        = & \sum_\zeta\frac{p_{k'}}{\prod_{l=1}^N p_{s'_l}}\tr\left(U_{SE}^\dag \Pi_{k'}\Pi_{\{s'\}}\Pi_{\{n'\}}U_{SE}\Pi_{\{s\}}\Pi_{\{n\}}\Pi_{k}\right)\prod_{j=1}^N p_{s_j}\prod_{m=1}^N p_{n_m} \nonumber \\
        = & \sum_{\zeta/\{k,n'\}}\frac{p_{k'}}{\prod_{l=1}^N p_{s'_l}}\tr\left(U_{SE}^\dag \Pi_{k'}\Pi_{\{s'\}}U_{SE}\Pi_{\{s\}}\Pi_{\{n\}}\right)\prod_{j=1}^N p_{s_j}\prod_{m=1}^N p_{n_m} \nonumber \\
        = & \sum_{\zeta/\{k,s,n,k',n'\}}\frac{1}{\prod_{l=1}^N p_{s'_l}}\tr\left(\rho'_S\Pi_{\{s'\}}U_{SE}\bigotimes_{j=1}^N\rho_{S_j}\rho_{E_j}U_{SE}^\dag \right) \nonumber \\
        = & \sum_{s'}\frac{1}{\prod_{l=1}^N p_{s'_l}}\tr\left(\rho'_S\Pi_{\{s'\}}\bigotimes_{j=1}^N\rho'_{S_j}\right) \nonumber \\
        = & \sum_{s'} \tr\left(\rho'_S\Pi_{\{s'\}}\right) \nonumber \\
        = & \tr\rho'_S \nonumber \\
        = & 1.
    \end{align}
    \end{widetext}
    The first to the second line is due to $\Pi_k$ and $\Pi_{\{n\}}$ are eigenvectors of $\rho_S$ and $\rho_E$, with the eigenvalues $p_k$ and $\prod_{m=1}^N p_{n_m}$. The second to the third line is to apply the completeness of $\Pi_k$ and $\Pi_{\{n'\}}$. The third to the fourth line is summing up the indexes $s,n,k'$, which gives 
    \begin{equation}
        \rho_{S_j} = \sum_{s_j}p_{s_j}\Pi_{s_j},\quad \rho_{E_j} = \sum_{n_j}p_{n_j}\Pi_{n_j},\quad \rho'_{S} = \sum_{k'}p_{k'}\Pi_{k'}.
    \end{equation}
    The fourth to the fifth line is from obtaining the final state from $U_{SE}$ and tracing out the environment. The fifth to the sixth line is from the decomposition $\rho'_{S_j} = \sum_{s'_j}p_{s'_j}\Pi_{s'_j}$. The sixth to the seventh line is because of the completeness of $\Pi_{\{s'\}}$.
\end{proof}
Although the quasiprobability $\mathcal Q^{\text F}$ can take complex or negative values, the Jensen-like equation 
\begin{equation}
    e^{-\langle \Delta\iota\rangle_{\mathcal Q^{\text F}[\zeta]}}\leq \langle e^{-\Delta\iota} \rangle_{\mathcal Q^\mathrm{F}[\zeta]}
\end{equation}
still holds, as the information inequality $\Delta\mathcal I=\langle \Delta\iota\rangle_{\mathcal Q^{\text F}[\zeta]}\geq 0$ is satisfied. Specifically, the real part of the quasiprobability must satisfy a recently established necessary and sufficient condition for Jensen’s inequality \cite{horvath2024necessary}. This provides a new perspective for studying the quasiprobability distribution in decoherence dynamics, a subject we will explore in a separate work.

Corresponding to the quasiprobability of the forward process $\mathcal Q^{\text F}$, the quasiprobability for the reversed process is defined as
\begin{equation}
\label{def:Q_B}
    \mathcal Q^\text{B}[\zeta] = \tr\left(\Pi_{k'}\Pi_{\{s'\}}\Pi_{\{n'\}}U_{SE}\Pi_{\{s\}}\Pi_{\{n\}}\Pi_{k}U_{SE}^\dag\rho'_S\rho^\text{ref}_E\right),
\end{equation}
where $\rho^\text{ref}_E = \bigotimes_{j=1}^N \rho'_{E_j}$ is the environmental reference state and $\rho'_{E_j}$ is the reduced density matrix from the final state $\rho'_{E}$. The symmetry between the forward and reversed quasiprobabilities then leads to the detailed fluctuation theorem. 
\begin{theorem}
    The stochastic total quantum correlation change $\Delta\iota$ satisfies the detailed fluctuation relation
    \begin{equation}
        \frac{\mathcal Q^\mathrm{F}[\zeta]}{\prod_{j=1}^N \mathcal P_{j}^\mathrm{F}[\xi_j]} e^{-\Delta\iota} = \frac{\mathcal Q^\mathrm{B}[\zeta]}{\prod_{j=1}^N \mathcal P_{j}^\mathrm{B}[\xi_j]}.
    \end{equation}
\end{theorem}
\begin{proof}
    First, we can rewrite the forward and backward quasiprobabilities defined in Eqs. (\ref{def:Q_F}) and (\ref{def:Q_B}) as
    \begin{subequations}
    \begin{multline}
        \mathcal Q^\text{F}[\zeta] \\
        =\tr\left(U_{SE}^\dag \Pi_{k'}\Pi_{\{s'\}}\Pi_{\{n'\}}U_{SE}\Pi_{\{s\}}\Pi_{\{n\}}\Pi_{k}\right)p_k \prod_{j=1}^N p_{n_j}, 
    \end{multline}
    \begin{multline}
        \mathcal Q^\text{B}[\zeta] \\
        =\tr\left(\Pi_{k'}\Pi_{\{s'\}}\Pi_{\{n'\}}U_{SE}\Pi_{\{s\}}\Pi_{\{n\}}\Pi_{k}U_{SE}^\dag\right)p_{k'}\prod_{j=1}^N p_{n'_j}.
    \end{multline}
    \end{subequations}
    Because of the cyclic of trace operation, we can easily see 
    \begin{equation}
    \label{eq:Q_F_over_Q_B}
        \frac{\mathcal Q^\text{F}[\zeta]}{\mathcal Q^\text{B}[\zeta]} = \frac{p_k\prod_{j=1}^N p_{n_j}}{p_{k'}\prod_{l=1}^N p_{n'_l}}.
    \end{equation}
    Besides, the probabilities of the local dynamics has the relation
    \begin{equation}
        \frac{\mathcal P_{j}^{\text F}[\xi_j]}{\mathcal P_{j}^{\text B}[\xi_j]} = \frac{p_{s_j}p_{n_j}}{p_{s'_j}p_{n'_j}}.
    \end{equation}
    Combining the above two relations, we get
    \begin{equation}
        \frac{\mathcal Q^\text{B}[\zeta]}{\mathcal Q^\text{F}[\zeta]}\frac{\prod_{j=1}^N \mathcal P_{j}^{\text F}[\xi_j]}{\prod_{l=1}^N \mathcal P_{l}^{\text B}[\xi_l]} = \frac{p_{k'}\prod_{j=1}^N p_{s_j}}{p_k\prod_{l=1}^N p_{s'_l}},
    \end{equation}
    which is identical to $e^{-\Delta\iota}$.
\end{proof}

The quasiprobability approach not only captures the essence of quantum statistics but also offers a powerful framework for extending the applicability of fluctuation theorems to quantum information dynamics. The integral fluctuation relation can be constructed from the quasiprobability because of the normalization of quasiprobability. Conceptually, our results imply that the Bayesian retrodiction based on classical probability can be extended to the quasiprobability regime \cite{buscemi2021fluctuation,aw2021fluctuation}.



\section{Fluctuation theorem of coherence dynamics}

\label{sec:quantum_coherence}

The fluctuation theorems for total classical and quantum correlations we established naturally imply a corresponding fluctuation theorem for coherence dynamics, due to the decomposition given by Eq. (\ref{eq:Delta_I_decomposition}), which holds when the final state $\rho'_S$ is completely dephased. The information inequality $\Delta\mathcal C\geq 0$ is also guaranteed. Explicitly, we define the stochastic coherence change as 
\begin{equation}
\label{eq:delta_c}
    \Delta c = \ln p_{k} - \ln p_{\{s\}} - \left(\ln p_{k'} - \ln p_{\{s'\}}\right),
\end{equation}
which is consistently related to $\Delta\iota_\text{cl}$ and $\Delta\iota$, namely
\begin{equation}
    \Delta\iota = \Delta\iota_\text{cl} + \Delta c.
\end{equation}
Its expectation value satisfies 
\begin{equation}
    \langle \Delta c\rangle_{\mathcal Q^{\text F}[\zeta]} = \Delta\mathcal C.
\end{equation}
Furthermore, the integral and detailed fluctuation theorems for $\Delta c$ can be established as follows. 
\begin{theorem}
    The stochastic coherence change $\Delta c$ satisfies the integral fluctuation relation
    \begin{equation}
    \label{eq:FT_c}
        \langle e^{-\Delta c} \rangle_{\mathcal Q^\mathrm{F}[\zeta]} = 1,
    \end{equation}
    and the detailed fluctuation relation
    \begin{equation}
        \frac{\mathcal Q^\mathrm{F}[\zeta]}{\mathcal P^\mathrm{F}[\xi]} e^{-\Delta c} = \frac{\mathcal Q^\mathrm{B}[\zeta]}{\mathcal P^\mathrm{B}[\xi]}.
    \end{equation}
\end{theorem}
\begin{proof}
    Similar as Eq. (\ref{proof:FT_q_mutual_information}), we have
    \begin{widetext}
        \begin{align}
        \langle e^{-\Delta c} \rangle_{\mathcal Q^{\text F}[\zeta]} = & \sum_\zeta\frac{p_{k'}p_{\{s\}}}{p_{k}p_{\{s'\}}} \tr\left(U_{SE}^\dag \Pi_{k'}\Pi_{\{s'\}}\Pi_{\{n'\}}U_{SE}\Pi_{\{s\}}\Pi_{\{n\}}\Pi_{k}\rho_S\rho_E\right) \nonumber \\
        = & \sum_\zeta\frac{p_{k'}}{p_{\{s'\}}}\tr\left(U_{SE}^\dag \Pi_{k'}\Pi_{\{s'\}}\Pi_{\{n'\}}U_{SE}\Pi_{\{s\}}\Pi_{\{n\}}\Pi_{k}\right)p_{\{s\}}\prod_{j=1}^N p_{n_j} \nonumber \\
        = & \sum_{\zeta/\{k,n'\}}\frac{p_{k'}}{p_{\{s'\}}}\tr\left(U_{SE}^\dag \Pi_{k'}\Pi_{\{s'\}}U_{SE}\Pi_{\{s\}}\Pi_{\{n\}}\right)p_{\{s\}}\prod_{j=1}^N p_{n_j} \nonumber \\
        = & \sum_{\zeta/\{k,s,n,k',n'\}}\frac{1}{p_{\{s'\}}}\tr\left(\rho'_S\Pi_{\{s'\}}U_{SE}\tilde\rho_{S}\rho_{E}U_{SE}^\dag \right) \nonumber \\
        = & \sum_{s'}\frac{1}{p_{\{s'\}}}\tr\left(\rho'_S\Pi_{\{s'\}}\tilde\rho'_S\right).
    \end{align}
    \end{widetext}
    Suppose that the final state $\rho_S'$ is a completely dephased state at the local eigenbasis satisfying Eq. (\ref{eq:dephasing_condition}), which gives $\rho'_S\Pi_{\{s'\}} = p_{\{s'\}} \Pi_{\{s'\}}$. Thus, we have
\begin{equation}
        \langle e^{-\Delta c} \rangle_{\mathcal Q^{\text F}[\zeta]} = \sum_{s'} \tr\left(\Pi_{\{s'\}}\tilde\rho'_S\right) = \tr\tilde\rho'_S = 1.
\end{equation}
The two-point measurement statistics have the relation
    \begin{equation}
        \frac{\mathcal P^\text{F}[\xi]}{\mathcal P^\text{B}[\xi]} = \frac{p_{\{s\}}\prod_{j=1}^N p_{n_j}}{p_{\{s'\}}\prod_{l=1}^N p_{n'_l}}.
    \end{equation}
    Combining Eq. (\ref{eq:Q_F_over_Q_B}), we have
    \begin{equation}
        \frac{\mathcal Q^\text{B}[\zeta]}{\mathcal Q^\text{F}[\zeta]} \frac{\mathcal P^\text{F}[\xi]}{\mathcal P^\text{B}[\xi]} = \frac{p_{k'}p_{\{s\}}}{p_kp_{\{s'\}}},
    \end{equation}
    which is identical to $e^{-\Delta c}$.
\end{proof}

Note that any two of the three detailed fluctuation relations for $\Delta\iota_\text{cl}$, $\Delta\iota$, and $\Delta c$ uniquely determine the third one, by virtue of the additive decomposition $\Delta\iota = \Delta\iota_\text{cl} + \Delta c$. 


\section{Three-qubit examples}

\label{sec:example}

We illustrate our results using a system of three qubits, each locally coupled to a one-qubit environment. The initial environment qubits are prepared as a product state, namely $\rho_E = \bigotimes_{j=1}^3\rho_{E_j}$. We randomly generate the initial three-qubit state $\rho_S$, the environment states $\rho_{E_j}$, and their interactions $U_{SE} = \bigotimes_{j=1}^3 U_{S_jE_j}$. When analyzing the coherence change, we set each local interaction to the two-qubit swap gate \cite{nielsenQuantumComputationQuantum2010}, $U_{S_jE_j} = \text{SWAP}$, to satisfy the completely dephasing condition given by Eq. (\ref{eq:dephasing_condition}). The corresponding code is publicly available on \cite{code}. As shown in Fig. \ref{fig_ft}, randomness in initial states leads to fluctuations in the changes of coherence and correlations, yet the integral fluctuation relations are always exactly satisfied.

\begin{figure}[t]
\includegraphics[width=1\columnwidth]{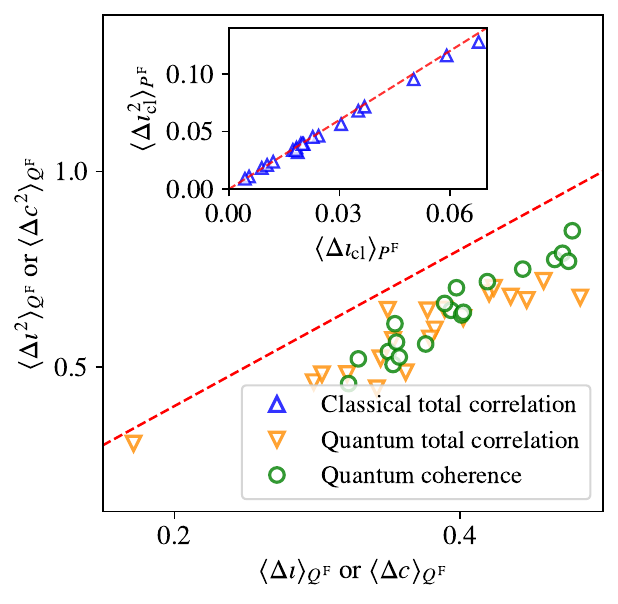}
\caption{The relationship between the expectation value and the second moment distribution of the total classical and quantum correlations, as well as the coherence change. The red dashed line represents $\langle x^2\rangle = 2\langle x\rangle$. The results are obtained from the data in Fig. \ref{fig_ft}.}
\label{fig_second_moment}
\end{figure}

We further verify that the second moment relation 
\begin{equation}
    \langle x^2\rangle = 2\langle x\rangle + \mathcal O(x^3),
\end{equation}
which follows from the fluctuation theorem $\langle e^{-x}\rangle = 1$. As shown in Fig. \ref{fig_second_moment}, the second moments of the stochastic total classical and quantum correlation changes, as well as the coherence change, are bounded by twice their expectation values. 



\section{Conclusions and outlooks}

\label{sec:conclusion}

We have formulated both integral and detailed fluctuation theorems that unify the dynamics of many-body coherence and correlations, using a quasiprobability framework to faithfully capture quantum features and the statistics of noncommuting observables. Not only the information inequalities are included in the information fluctuation theorems, but also providing a quantitative statistical description of quantum information dynamics. We further anticipate that the quasiprobability approach can be extended to a wide range of quantum resources, including quantum discord \cite{Ollivier2001QuantumDA}, quantum steering \cite{Uola2019QuantumS}, quantum imaginarity \cite{Hickey2018QuantifyingTI,Xu2024CoherenceAI}, and entanglement asymmetry \cite{Ares2022EntanglementAA,Ares2025TheQM}. If we interpret each local unitary evolution $U_{S_jE_j}$ as a process of work extraction \cite{Skrzypczyk2013WorkEA}, we can further explore the thermodynamic significance of our results. It would also be interesting to investigate the corresponding "thermodynamic" uncertainty relations based on our information fluctuation theorems \cite{Hasegawa2019FluctuationTU,Timpanaro2019ThermodynamicUR}. We leave these questions for future study.

We conclude this paper by some remarks on experimental verifying our established fluctuation theorems. Quantum fluctuation theorems have been experimentally verified on various platforms \cite{batalhao2014experimental,an2015experimental,cerisola2017using,masuyama2018information,Micadei2020ExperimentalVO,hernandez2021experimental,solfanelli2021experimental,hahn2023quantum,yan2024experimental,Li2025ExperimentalDO}. The fluctuation theorem for total classical correlation is based on the standard two-point measurement scheme, making its experimental verification straightforward. Verifying the fluctuation theorem for total quantum correlation or coherence dynamics requires measuring the stochastic quantum correlation change $\Delta\iota$ defined in Eq. (\ref{eq:delta_iota_q}) or coherence change $\Delta c$ defined in Eq. (\ref{eq:delta_c}) and the quasiprobability $\mathcal Q^\text{F}$ defined in Eq. (\ref{def:Q_F}). The former requires joint state measurements, which are experimentally feasible for several qubits. The latter is not directly measurable, but can be inferred through the weak two-point measurement scheme \cite{johansen2007quantum,hernandez2024projective} or the interferometric scheme \cite{Dorner2013ExtractingQW,Lostaglio2022KirkwoodDiracQA,HernandezGomez2024InterferometryOQ}. In the weak measurement scheme, non-orthogonal operators are applied, and the quasiprobability can be constructed from the combination of weak and projective measurement results. The interferometric scheme employs an auxiliary system to create a superposition of noncommutative projectors, allowing the quasiprobability to be extrapolated. This scheme shares the same principle as the Hadamard test used in quantum computing. Recently, approaches for determining quasiprobabilities using quantum circuits have been extensively studied \cite{wagner2024quantum}. Given the current capabilities of quantum computers, we anticipate that they will be able to verify our fluctuation theorems involving around 10 qubits.




\begin{acknowledgments}
KZ, XHW, and HLS thank the support by the NSFC (No. 12305028, No. 12275215, No. 12247103, No. 92365202, No. 12134015, and No. 12234019), National Key Research and Development Program of China (2024YFA0919600), Shaanxi Fundamental Science Research Project for Mathematics and Physics (Grant No. 22JSZ005), the Youth Innovation Team of Shaanxi Universities and Quantum Science, and Technology-National Science and Technology Major Project  under grant No. 2023ZD0300400. KZ is supported by the China Postdoctoral Science Foundation under Grant No. 2025M773421, Shaanxi Province Postdoctoral Science Foundation under Grant No. 2025BSHYDZZ017, and Scientific Research Program Funded by Education Department of Shaanxi Provincial Government (Program No.24JP186). HLS was supported by the Horizon Europe programme HORIZONCL4-2022-QUANTUM-02-SGA via Project No. 101113690 (PASQuanS2.1). 
\end{acknowledgments}

\section*{DATA AVAILABILITY}

The data that support the findings of this article are openly available at \cite{code}, embargo periods may apply.






\providecommand{\noopsort}[1]{}\providecommand{\singleletter}[1]{#1}%

\end{document}